\documentclass[letterpaper,twocolumn,10pt]{article}
\usepackage{usenix-2020-09}

\newif\iffullversion\fullversiontrue

\usepackage{tikz}
\usepackage{amsmath}
\usepackage{graphicx}
\usepackage{tabularx}
\usepackage{booktabs}
\usepackage{cite}
\usepackage{microtype}
\usepackage{paralist}
\usepackage{xspace}
\usepackage{amsfonts}
\usepackage{amsmath}
\usepackage{mathtools}
\usepackage{algorithm}
\usepackage[noend]{algpseudocode}
\usepackage{enumitem}
\usepackage{tikz}
\usepackage[singlelinecheck=false]{caption}
\usetikzlibrary{calc}
\usepackage{varwidth}
\usepackage{amsthm}

\usepackage{color}
\definecolor{scommentColor}{rgb}{0,0.6,0.8}
\definecolor{acommentColor}{rgb}{0.8,0.3,0}
\definecolor{mcommentColor}{RGB}{16,109,28}
\definecolor{jcommentColor}{RGB}{150,0,150}
\newif \ifcomments \commentsfalse
\ifcomments
\newcommand{\sunoo}[1]{{\color{scommentColor} /* S: #1 */}}
\newcommand{\andres}[1]{{\color{acommentColor} /* A: #1 */}}
\newcommand{\mike}[1]{{\color{mcommentColor} /* M: #1 */}}
\newcommand{\jack}[1]{{\color{jcommentColor} /* J: #1 */}}
\else
\newcommand{\sunoo}[1]{}
\newcommand{\andres}[1]{}
\newcommand{\mike}[1]{}
\newcommand{\jack}[1]{}
\fi

\newif \ifaftersub \aftersubtrue

\definecolor{newdiffcolor}{rgb}{0, 0.66, 0.47}
\definecolor{newdiffcolor}{rgb}{0, 0, 0}
\newcommand{\newdiff}[1]{{\color{newdiffcolor} #1}}

\newcommand{\add}{\texttt{add}}
\newcommand{\update}{\texttt{update}}
\newcommand{\delete}{\texttt{deregister}}

\newcommand{\Modules}{{\sf Modules}\xspace}
\newcommand{\Reg}{{\sf Register}\xspace}
\newcommand{\UpdateReg}{{\sf UpdateRegistration}\xspace}

\newcommand{\Maintenance}{{\sf Maintenance}\xspace}
\newcommand{\Oversight}{{\sf Oversight}\xspace}

\newcommand{\Query}{{\sf Query}\xspace}
\newcommand{\Audit}{{\sf Audit}\xspace}

\newcommand{\subh}[1]{\par \noindent \textit{{#1}}.}
\renewcommand{\paragraph}[1]{\subh{#1}}

\newcommand{\reg}{\mathsf{R}}
\newcommand{\regcom}{\mathsf{com}}
\newcommand{\regupdate}{\mathsf{R.Update}}
\newcommand{\reglookup}{\mathsf{R.Lookup}}
\newcommand{\regverlookup}{\mathsf{VerLookup}}
\newcommand{\reghistory}{\mathsf{R.History}}
\newcommand{\regverhistory}{\mathsf{VerHistory}}
\newcommand{\regaudit}{\mathsf{Audit}}

\newcommand{\Enc}{\mathsf{Enc}}
\newcommand{\Dec}{\mathsf{Dec}}
\newcommand{\KDF}{\mathsf{KDF}}
\newcommand{\advA}{\mathcal{A}}

\newcommand{\encode}{{\textsf{encode}}}
\newcommand{\match}{{\textsf{match}}}

\newcommand{\voters}{{\cal V}}
\newcommand{\voter}{\ensuremath{V}}

\newcommand{\thirdparties}{{\cal T}}
\newcommand{\thirdparty}{\ensuremath{T}}
\newcommand{\vrdb}{\mathsf{D}}
\newcommand{\vrdbcols}{{\cal C}}
\newcommand{\vrdbcol}{\ensuremath{C}}
\newcommand{\bulletin}{\mathsf{B}}
\newcommand{\pool}{\mathsf{Q}}
\newcommand{\id}{\mathsf{ID}}
\newcommand{\access}{{\textsf{access}}}
\newcommand{\public}{{\textsf{public}}}
\newcommand{\updrec}{{\textsf{updrec}}}
\newcommand{\concat}{\:\|\:}
\newcommand{\false}{\mathsf{false}}
\newcommand{\true}{\mathsf{true}}
\newcommand{\basesys}{\textsf{S}\xspace}
\newcommand{\sysname}{\textsf{VRLog}\xspace}
\newcommand{\sysnamepriv}{\textsf{VRLog$^\times$}\xspace}
\newcommand{\assignRand}{{\leftarrow}\vcenter{\hbox{\tiny\rmfamily\upshape\$}}}

\algrenewcommand{\algorithmiccomment}[1]{\footnotesize\hskip3em$//$ #1}
\newtheorem{thm}{Theorem}[section]

\renewcommand{\paragraph}[1]{\smallskip\noindent\textbf{#1}\enskip}

\graphicspath{ {./figures/} }

\newlength{\saveparindent}
\newlength{\saveparskip}
\newcounter{ctr}

\newenvironment{newitemize}{%
\begin{list}{\mbox{}\hspace{5pt}$\bullet$\hfill}{\labelwidth=15pt%
\labelsep=5pt \leftmargin=20pt \topsep=3pt%
\setlength{\listparindent}{\saveparindent}%
\setlength{\parsep}{\saveparskip}%
\setlength{\itemsep}{3pt} }}{\end{list}}

\newenvironment{newenum}{%
\begin{list}{{\rm (\arabic{ctr})}\hfill}{\usecounter{ctr} \labelwidth=17pt%
\labelsep=5pt \leftmargin=22pt \topsep=3pt%
\setlength{\listparindent}{\saveparindent}%
\setlength{\parsep}{\saveparskip}%
\setlength{\itemsep}{2pt} }}{\end{list}}

\theoremstyle{definition}

\newenvironment{pfsketch}
{\begin{proof}[Proof (sketch)]}
{\end{proof}}

\begin{document}

\date{}

\title{Cryptographic Verifiability for Voter Registration Systems}

\author{
{\rm Andr\'{e}s F\'{a}brega\thanks{Joint first authors.}}\\
Cornell University
\and
{\rm Jack Cable$^*$}\\
Corridor
\and
{\rm Michael A. Specter}\\
Georgia Institute of Technology
\and
{\rm Sunoo Park}\\
New York University
}

\maketitle

\begin{abstract}
  \newdiff{
Voter registration systems are a critical---and surprisingly understudied---element of most high-stakes elections. Despite a history of targeting by adversaries, relatively little academic work has been done to increase visibility into how voter registration systems keep voters' data secure, accurate, and up to date. 
Enhancing transparency and verifiability could help election officials and the public detect and mitigate risks 
to this essential component of electoral processes worldwide.

This work introduces \emph{cryptographic verifiability} for voter registration systems.
Based on consultation with diverse expert stakeholders that support elections systems,
we precisely define the requirements for cryptographic verifiability in voter registration
and systematize the practical challenges that must be overcome for near-term deployment.

We then introduce \sysname}, the first system to bring strong verifiability to voter registration. \sysname enables election officials to provide a transparent log that (1) allows voters to verify that their registration data has not been tampered with and (2) allows the public to monitor update patterns and database consistency.
We also introduce \sysnamepriv, an enhancement to \sysname that offers cryptographic privacy to voter deduplication between jurisdictions---a common maintenance task currently performed in plaintext or using trusted third parties. Our designs rely on standard, efficient cryptographic primitives, and are backward compatible with existing voter registration systems. 
Finally, we provide an open-source implementation of \sysname and benchmarks to demonstrate that the system is practical---capable of running on low-cost commodity hardware and scaling to support databases the size of the largest U.S. state voter registration systems.

\end{abstract}

\section{Introduction}\label{sec:intro}

Democracies rely on \textit{voter registration systems}~\cite{pew-countries} to maintain an up-to-date list of who is eligible to vote, to scale to potentially thousands of jurisdictions and millions of voters, and to protect voters' sensitive personal information.\footnote{Exactly what voter data is made available to the public depends heavily on the jurisdiction's individual laws and procedures.} 
These systems serve a foundational role in elections: for most voters, the registration system ultimately determines who is allowed to cast a ballot.\footnote{In many states in the U.S., same day voter registration allows unregistered voters to register and cast a ballot on election day. Furthermore, the Help America Vote Act of 2002 requires all U.S. states with voter registration systems to allow voters to cast a provisional ballot should they contest their name not appearing on the voter registration list~\cite{hava}.}
As a result, trust in the security of the voter registration system is a necessary prerequisite for confidence in the election system as a whole, and undermining voter registration can be a high-value target for malicious actors.

Integrity and secrecy of voter registration systems have been the source of concern for some time. 
Claims of insider threats altering voter rolls have repeatedly reached the popular press --- including both registration of ineligible voters~\cite{seelyeVoterFraudNew2017} and removal of legitimate voters~\cite{GeorgiaOfficialsSeek2021}.
Attacks against voter registration systems are well documented: there are credible reports of Russian and Iranian government-affiliated actors compromising U.S. state voter registration databases in both the 2016 and 2020 U.S. elections~\cite{ssci,alaska}. While there is no evidence that malicious actors have ever modified voter records in the U.S. at scale, both spurious claims and real attacks can harm the perceived legitimacy of the democratic process.

\newdiff{In light of such concerns, transparency in voter registration systems is critical to bolster election security and public confidence. Many jurisdictions worldwide, including nearly all U.S. states, take steps to provide some transparency into their systems, such as offering
parts of their voter rolls for public inspection. Existing measures like publishing voter rolls, though valuable, only provide a narrow form of transparency. Cryptographic verifiability provides many guarantees beyond existing transparency measures---in some cases, offering guarantees that would not be possible absent cryptography.

\begin{table*}[ht]
\centering\small
\renewcommand{\arraystretch}{1.5}

\begin{tabularx}{\linewidth}{lX}

\hline
\textbf{Stakeholder} & 
\textbf{Can verify the following system data and properties} \\
\hline
Election officials & (1) That any modifications to the voter registration database will be logged; and (2) any unauthorized modifications will be easy to detect and revert.
\\ 
Privileged auditors & That any inconsistencies in the
voter data they receive from election officials would be detectable by voters themselves based on public information.
\\ 
Registered voters & (1) Their current registration status; (2) whether
their registration data was recorded accurately; and (3) if and when their registration status or data had been modified.
\\
The public & (1) Whether any given version of the voter registration database is consistent with the public record of all updates to the voter registration database, including registrations, modifications, and maintenance operations; (2) whether that public record has been consistent over time; and (3) that the public record is append only.
\\ \hline
\end{tabularx}

\caption{\small Informal description of the verifiability properties that are achievable with a verifiable voter registration system. We provide formal descriptions of these guarantees (\S\ref{sec:voter-reg-background}), and develop a prototype to demonstrate that such a system is practical (\S\ref{sec:impl}).}
\label{tbl:intro}
\end{table*}

In this work, we initiate the study of \emph{cryptographic verifiability for voter registration systems}. We provide a threat model and precisely define the security requirements that a cryptographically \emph{verifiable voter registration system} should provably satisfy. As summarized in Table~\ref{tbl:intro}, verifiable voter registration systems enable different stakeholders to verify properties of the system and data, such as the integrity of (their) data, history of modification, and maintenance processes---all while guaranteeing cryptographic privacy protection for the sensitive voter information in the database.}

\newdiff{We justify the need and relevancy of these guarantees for election security, which are motivated by real-world threats to registration systems. Critically, these guarantees should rely on minimal trust assumptions, and should hold even if significant election infrastructure has somehow been compromised. Cryptographic verifiability (1) helps voters better monitor their own records, (2) allows the public, including civil society and advocacy groups, to continuously verify various aspects of voter registration and maintenance, and (3) 
provides election officials additional evidence to counter illegitimate claims of registration compromise.

\paragraph{Practical deployment challenges.} Due to the complexities, high stakes, and practical administration of elections, modifications to election infrastructure entail strong constraints, the understanding of which is critical for real-world adoption of new systems. Therefore, in order to define the security goals for strong registration verifiability and to understand the practical challenges and considerations for deploying a new voter registration system, we consulted with diverse stakeholders in the U.S., including current and former election officials, civil society organizations, and practitioners. From these conversations, we \emph{systematize the practical requirements} for new registration systems, and distill a list of design goals.

We identify four essential needs from our stakeholder consultations, as further elaborated in Section~\ref{sec:stakeholder-constraints}:} (1) interfacing with pre-existing analog processes; (2) interfacing with pre-existing digital processes; (3) designing agnostic to voters' digital literacy; and (4) accommodate jurisdictional variations in how data is stored and protected.

\newdiff{\paragraph{\sysname: a practical prototype.} Having introduced the security requirements and practical constraints, we then propose \sysname, the first system to bring cryptographic verifiability to voter registration systems. As our stakeholder consultations showed, backward compatibility and straightforward system integration
are essential for near-term deployment of a new registration system. Thus, \sysname assembles a deliberately simple set of cryptographic tools to provide strong registration verifiability. In a nutshell,} \sysname augments existing voter registration systems with a cryptographically verifiable public log, which records all voter registration activity (e.g., registrations, maintenance operations, and deletions) while encrypting sensitive voter information such that only those authorized to access an individual voter record (or specific fields thereof) can do so. As a result, patterns of activity are publicly visible---for example, unusually large-scale modifications would be publicly evident---and any irregularity in encrypted data can be traced through a cryptographically binding audit trail. We prove that \sysname satisfies strong verifiability properties and show it has acceptable speed and storage overhead at the scale of the largest U.S. state voter registration systems.

The key cryptographic tool underlying \sysname is a 
\emph{cryptographically verifiable registry}.
\sysname presents a novel application of
verifiable registries, which have to date seen impact largely
\emph{within} cryptography application contexts, such as key/certificate transparency~\cite{trillian,certificatetransparency}.
Yet verifiable registries are a versatile tool
with potential impact well beyond cryptography contexts;
\sysname is among the first few 
practical examples with broader applications.

Prior work applying verifiable registries to real-world use-cases such as key transparency, may at first appear to provide a straightforward solution to our problem (see Section~\ref{sec:related-work}). However, voter registration systems have practical constraints unique to the elections context that make an out-of-the-box solution unworkable, related to the four constraints highlighted by stakeholders, mentioned above. Each of these constraints informed our tailored solution in \sysname (see Section~\ref{sec:stakeholder-constraints}), and some of our techniques may be of independent interest for similar challenges in other registry applications.

\paragraph{\sysnamepriv: a verifiable privacy-enhancing extension.}
As an extension to \sysname, we introduce \sysnamepriv, which verifiably enhances the privacy of voter registration systems when \emph{interfacing with external data sources for maintenance}. Currently, routine maintenance processes for keeping voter information up-to-date---such as comparing with another state's voter registration database to identify duplicates---either rely on a trusted third party intermediary or use plaintext data exchange. We enhance \sysname to support \emph{privacy-preserving record linkage (PPRL)} protocols that reveal significantly less information to third-party entities that interact with subsets of voter data for maintenance purposes. Importantly, we introduce an additional cryptographic step to enable our registry to \emph{verifiably} interface with PPRL protocols from prior work.

\paragraph{Implementation.}
We implement \sysname on top of the Trillian project, which provides a general framework for building transparency logs~\cite{trillian}. Our \href{https://github.com/cablej/vrlog}{open source implementation}, which comprises over 1,000 lines of code, is available to the community and includes functionality to register voters, encrypt fields, update voter info, get voters, prove and verify membership, and prove and verify the append-only nature of the list. We demonstrate that \sysname can efficiently support even the largest voter registration databases with tens of millions of records on commodity hardware. 

\paragraph{Summary of contributions.} We initiate the study of cryptographic verifiability for voter registration systems.
\begin{newitemize}
    \item We define the \emph{security goals and threat models} for cryptographic registration verifiability (\S~\ref{sec:voter-registration-verifiability}), motivated by threats and attacks on registration systems.
    \item We systematize the \emph{practical considerations and requirements} necessary for near-term deployment of such a system, guided by stakeholder consultations (\S~\ref{sec:voter-reg-background}).
    \item We present \sysname (\S~\ref{sec:vrlog}), a new verifiable voter registration system with provable security, which is compatible with the practical considerations we identified.
    \item We introduce an extension, \sysnamepriv (\S~\ref{sec:privacy-ehancing-extensions}), which further enhances \emph{transparency} and \emph{confidentiality} in routine maintenance protocols in which U.S. states currently exchange sensitive voter data in the clear.
    
    \item We provide an efficient implementation of \sysname (Section~\ref{sec:impl}), and provide detailed performance estimates for \sysnamepriv (Section~\ref{subsec:privacy-extension-framework}), demonstrating the immediate practicality of the respective system designs.
\end{newitemize}

\section{Related Work}

To our knowledge, ours is the first work that brings \emph{cryptographic transparency} guarantees to voter registration systems. We highlight a few related lines of work below. 

\label{sec:related-work}
\paragraph{Transparency in voting systems.} A substantial body of work has focused on the transparency of voting processes other than registration: notably, casting and tallying~\cite{adida2008helios,bell2013star,benaloh2015end} and post-election auditing~\cite{lindeman2012gentle,benaloh2011soba,lindeman2012bravo}. Voter registration is outside the scope of these works, which are tailored to the security requirements of casting ballots and processing them after casting. While this line of work and ours share the common use of a public bulletin board, the substance of our constructions differs almost entirely. Casting, tallying, and auditing each have unique threat models and security requirements (e.g.,  ballot secrecy and coercion resistance) that are incomparable to the unique requirements of registration (e.g., verifying continued accuracy over long time periods, and cross-database checks).\footnote{See Section~\ref{sec:voter-reg-background} for a full description of voter registration requirements.}

\paragraph{Voter registration security.} Academic work on voter registration systems is relatively sparse. We build upon a recent work by Cable et al.~\cite{cable2023systematization}, which introduces a definitional framework that models voter registration systems (see Section~\ref{sec:voter-reg-background}). Limited prior work on cryptography for voter registration considers issues \textit{other} than verifiability. For instance, Merino et al. present a system that allows voters to create indistinguishable real and fake credentials to mitigate coercion concerns when using an untrusted device to register to vote~\cite{merino2022trip}. Another line of work attempts to detect anomalies in public voter registration lists using statistical models~\cite{voteshield2,cao2022bayesian,alvarez2009interstate,ansolabehere2010quality,shino2020verifying}. Some have proposed using blockchains to store voters' data~\cite{panja2021secure}, though only as part of a broader design for an \emph{electronic-only} voting system. However, such proposals do not support important (and often legally required) aspects of voter registration such as list maintenance processes and non-electronic registration, and publicly reveal substantially more voter information than current systems \cite{panja2021secure}.

\paragraph{Real-world list maintenance.} Some U.S. states have adopted a trusted-party-based approach to cross-check their voter lists. 
The most notable initiative is the non-profit Electronic Registration Information Center (ERIC)~\cite{eric}, which provides a valuable resource to aid states in voter list maintenance. While ERIC uses cryptographic hashing, it still involves the transfer of more sensitive voter information than mathematically necessary using cryptographic techniques. 
Moreover, participation in ERIC is limited to less than half of U.S. states, with some states instead opting for bilateral exchanges.

\paragraph{Verifiable registries.} There is a rich literature on cryptographic construction of publicly verifiable data structures such as append-only logs~\cite{chase2019seemless,meiklejohn2020think,tomescu2019transparency,melara2015coniks,tyagi2022versa,leung2022aardvark,crosby2009efficient}, which offer a range of different trade-offs between security, performance, and usability. 
Our work builds on Trillian, an implementation of a Merkle tree-based log~\cite{trillian}, which is both widely supported and provided adequate flexibility for \sysname. 

\paragraph{Applications of transparent data structures.} 
To date, most applications of transparent data structures such as verifiable registries are in the context of key transparency ~\cite{malvai2023parakeet,chase2019seemless,len2023optiks,melara2015coniks}, particularly for end-to-end encrypted messaging and certificate transparency. A handful of other applications 
focus on security of software and file systems: e.g., software supply chain security~\cite{newman2022sigstore, merrill2023speranza}, audit trails for versioning file systems~\cite{peterson2007design}, and binary transparency~\cite{al2018contour}. 
Yet verifiable registries have potential for impact well beyond these highly technical domains. \sysname is among a first few 
practical examples with broader application;    another example is a recent proposal for verifiable gun registries \cite{kamara2021gun}.

\section{Overview of Voter Registration Systems}\label{sec:voter-reg-background}
\newdiff{We provide background on voter registration systems in this section, highlighting key requirements and challenges.}

\newdiff{
\paragraph{Methodology.} To build our understanding of voter registration, we surveyed \emph{extensive public documentation}, encompassing the registration systems of all fifty U.S states\footnote{Except North Dakota, which does not have voter registration.}~\cite{ncsl_2019, ncsl_maintence, ncsl_avr, ncsl_sdr, brennan, Brennan06, Brennan09a, Brennan10, Brennan15, EAC05} and numerous other countries~\cite{Brennan09b,ACE-VR}.}

\newdiff{We complemented this information with \emph{informal, unstructured consultations with 16 diverse expert stakeholders}, including current and former election officials (9), other government roles (3), civil society organizations (3), and other practitioners (5).\footnote{These numbers do not sum to 16 as expertise was overlapping.} Most were U.S.-based, covering five states and federal government experience. See Appendix~\ref{a:interviews-table} for brief anonymized profiles. Given the nature of the consultations, this research did not require institutional review.\footnote{Our consultations do not constitute human subjects research for the purposes of applicable institutional review requirements because the stakeholders we spoke with were not research subjects, but rather expert consultants, and we did not collect identifiable information about them. 20 C.F.R. \S431. That is, we consulted them about their area of expertise, not about themselves.}

Our focus was the practical challenges of deploying (modifications to) voter registration systems, and identifying significant threats and concerns regarding their functionality.
We keep the stakeholders we consulted with anonymous, in accordance with their preferences, and to encourage candid discussion of sensitive topics related to election administration. }

\paragraph{Overview.} 
\iffullversion
    Voter registration systems serve many purposes, most notably to maintain an accurate list of registered voters in order to verify their eligibility on election day. At the heart of a voter registration system is the \emph{voter registration database (VRDB)}, where voter information is stored. At a high level, the process of voter registration consists of populating and maintaining the VRDB when voters register or update their data. In practice, VRDBs are subject to complex legal and operational requirements.
\fi

\iffullversion
    A first challenge is that voter registration systems must support and coordinate a wide array of registration methods, such as in person, email, fax, and web form. Data from these sources may be routed through intermediaries --- some outside of the election system, such as driving license agencies --- before being processed and combined in the VRDB. Ultimately, this list of registered, eligible voters must be available to election officials during elections. Elections may happen frequently---or even in parallel---and may impose additional restraints on the VRDB's functionality, such as freezing the modification of voter records for some period of time.
\fi

In addition, election officials must follow relevant laws and policies to ensure that voter records are kept up-to-date. Ideally, voters update their information when updates occur (e.g., if they move), but in practice many voters fail to do so. 
Election officials must therefore perform a variety of complex \emph{list maintenance} operations to persistently audit and update VRDB information~\cite{cable2023systematization}. This may involve interfacing with third parties (e.g., postal and social security services) to flag possible changes. In the U.S., states often compare their voter records with other states to identify duplicate voters.

Maintaining accurate voter information is challenging in part due to the fact that comparing personal data records is in itself a difficult task: two database records that refer to the same individual may appear very different (and, conversely, different individuals may share names, dates of birth, etc.). This is called a \emph{record linkage (RL)} problem (also known as \emph{entity resolution (ER})~\cite{binette2022almost,christophides2020overview}). Jurisdictions often use RL techniques to accurately maintain their data. In the U.S., for example, the nonprofit ERIC helps states detect duplicate voters by processing their data through a RL engine.

Election officials must establish a complex access control system for voter records, since many different entities may be approved to receive different subsets of the voter data based on applicable law and policies. In addition, the VRDB may need to support a variety of transparency requirements, which grant access to members of the public (such as voters, political candidates, and auditors) to review a subset of the VRDB; this further increases the required granularity of access control.

Additionally, voter registration systems must conform with the jurisdiction's legal requirements. For example, they are subject to important accessibility requirements, as any eligible citizen --- regardless of technical proficiency --- must be able to register. In addition, election officials may be required to protect voters who may be at risk if their data is published, such as survivors of intimate-partner violence, people with stalkers, or elected officials. This is often known as an Address Confidentiality Program in the U.S.

\paragraph{Stakeholder constraints.}
\label{sec:stakeholder-constraints}
We identify four key themes that featured consistently across our stakeholder consultations, indicating practical constraints specific to the electoral context: (1) the need to interface with pre-existing analog voter registration processes; (2) the need to interface with pre-existing digital voter registration processes; (3) the need for features to be accessible regardless of a voter's digital literacy; and (4) the need to accommodate myriad jurisdictional variations in how data is stored and protected (often as required by law). 

Our consultations highlighted that at least in the U.S., election officials' resource constraints~\cite{electionresources} can be severely limiting in the adoption of new initiatives, and strongly favor incremental change.
Moreover, they are generalists: election systems are just one aspect of election administration within their remit. Prioritization of time-sensitive issues like upcoming elections can cause delays to longer-term initiatives.
Our consultations also indicate a keen and justified awareness among election officials of the risks of changing critical infrastructure without adequate circumspection; combined with resource constraints, this further favors incremental change.
Finally, another important incentive for incremental change and backward compatibility is to maintain systems' accessibility, both to meet legal requirements and to serve a diverse electorate from whom digital literacy cannot be assumed, and for whom each system overhaul will come with confusion, delays, and other costs that election officials directly bear.

Our four key themes present challenges not present in prior work, such as key transparency: (1) analog processes are not involved;
(2) replacing an existing (centralized) digital system with a new one is more practical with fewer legal constraints and users have some baseline digital literacy;
(3) it is acceptable for security and verifiability guarantees to hold 
assuming interested parties have some technically competence;
and (4) the adoption of a single protocol and agreement on
which data to protect by all participating parties is often a preferred and feasible outcome.\footnote{E.g., in the U.S., the Constitution gives states the authority to govern the conduct of elections, so the adoption of a single protocol is not feasible.}
\newdiff{We thus had to build tailored solutions for these hurdles into \sysname, resulting in
an enhanced verifiable registry designed for 
the specific challenges and security needs of voter registration. These refinements are simple by design, and may be of independent interest for similar challenges in other registry applications.}

\paragraph{Voter registration modules.}\label{subsec:voter-reg-framework}
A challenge with defining protocols related to voter registration is that systems vary significantly between jurisdictions, and thus solutions must be adaptable to jurisdictional variations. Our model builds upon Cable et al's.~\cite{cable2023systematization} definitional framework for voter registration systems, which captures principles common to these systems, with flexible parameters to cover jurisdictional differences.

In this framework, the core functionality of a voter registration system is comprised of a series of \emph{modules}, which represent the basic processes and workflows of voter registration systems. The most relevant to our work are
\iffullversion
the following:
\begin{newitemize}
\item \emph{\Reg}, which receives as input an voter's data, and adds the voter's record to the VRDB if they are eligible.

\item \emph{\UpdateReg}, which receives as input an update from a registered voter, and updates their record in the VRDB.

\item \emph{\Maintenance}, which involves updates of voter data by election officials as part of their list maintenance processes, to correct out-of-date or inaccurate voter records. This may involve interaction with third party \emph{maintenance entities}~\cite{cable2023systematization}.

\item \emph{\Oversight}, which represents members of the public or pre-approved entities---which Cable et al. define as \emph{oversight entities}~\cite{cable2023systematization}---inspecting voter records to identify anomalies (e.g., voters who were incorrectly marked inactive), and reporting this to either the public or election officials.
\end{newitemize}
\fi

The voter registration systems of specific jurisdictions can be considered \emph{implementations} of these four high-level workflows. To capture the jurisdiction-specific details that vary across implementations, Cable et al. define \emph{jurisdictional parameters} and \emph{security policies}, which parameterize the modules (such as an access control policy, or a data change control policy), and define the security requirements of voter registration systems in terms of the \emph{completeness}, \emph{soundness}, and \emph{secrecy} of each of the four core modules,\footnote{We refer readers to Cable et al.'s works for a detailed breakdown of these properties and how they relate to each module.} which together comprise the security guarantees of the broader system.

\section{Voter Registration Verifiability}\label{sec:voter-registration-verifiability}

\subsection{Security Requirements}\label{subsec:security-requirements}
We first explain the security requirements that a verifiable registration system should (provably) satisfy. 
\iffullversion
    Like any security definition, this consists of (1) defining the desired security goals, and (2) specifying the conditions (i.e., threat model) under which these goals should be achieved. 
\fi

\paragraph{Verifiability goals.} A \emph{verifiable voter registration system} should provide four core verifiability guarantees, as summarized in Table~\ref{tbl:intro}: informally, (1) for \emph{voters} to be able to detect if their individual registration data has been modified or deleted without their knowledge; (2) for \emph{third parties}
(e.g., oversight and maintenance entities) 
to be able to detect if they received incorrect voter data from the election officials; (3) for the \emph{general public} to be able to observe patterns in modifications to voter records (though not the sensitive content of individual modifications); and (4) for \emph{election officials} to be able to detect modifications to voter data.
We formalize these verifiability requirements in Section~\ref{sec:security-analysis}.

\paragraph{Threat Model.}
There are six main types of entity that participate in \sysname (corresponding to the entity categories in Cable et al.'s framework): voters, election officials, the VRDB, maintenance entities, oversight entities, and the public. 

Voters trust election officials to not share their data except as allowed by the jurisdiction's access control policy. Under our threat model, voters do not need trust that their data has not been modified or deleted; rather, they can receive credible evidence that their registered data is in its correct state.
Similarly, maintenance entities and oversight entities do not need to trust that they have been provided with the correct data of the voters they are approved to access, and instead can receive proof that the data they receive is consistent.

We assume that any party, including voters, maintenance entities, oversight entities, and the general public may attempt to infer data that they are not approved to access. Our system should enable evidence-based verification of (the absence of) breaches of confidentiality or integrity of the VRDB.

\newdiff{These verifiability goals and threat model are motivated by notable concerns regarding voter registration systems. Prominent examples include \emph{voter purges} (removal of legitimate voters from the VRDB), \emph{voter data manipulation} (surreptitious modification of a voter's data), and \emph{registration stuffing} (adding illegitimate voters to the VRDB). These attacks can arise from compromise of election authorities, or impersonation of voters. Our verifiability goals and threat model are tailored to detect such attacks, in addition to accidental, but incorrect, modifications to voter data. Note that cryptographic registration verifiability does not intend to prevent or facilitate recovery from attacks on and errors in registration data. Rather, the goal is to make these \emph{detectable}, providing a route for voters to remedy their registration status if there are problems, or providing assurance that their information is correct and the registration system is functioning without problems.

}

\subsection{Design Goals}\label{subsec:design-goals}

Next, we outline a set of goals that further constrain the design of voter registration systems. These goals are motivated by the requirements and considerations outlined in Section~\ref{sec:voter-reg-background}. 

\paragraph{1. Backward compatibility.} To facilitate adoption, we argue that it must be straightforward to adopt a cryptographic verifiability system as an \emph{addition} to an existing voter registration system. 
The goal of a verifiability system is not to \emph{replace} a jurisdiction's VRDB, but rather to \emph{enhance} it with additional verifiability guarantees. This contrasts with verifiability constructions in other domains, such as key transparency~\cite{malvai2023parakeet,len2023optiks}, and is strongly motivated by practical concerns.

\newdiff{First, VRDBs are often part of complex election management systems, and have multiple use-cases and legal requirements that are orthogonal to our verifiability guarantees (e.g., configuring ballot layouts, redistricting, overseas/military balloting, etc). Election officials should be able to augment their existing voter registration systems with additional security guarantees without sacrificing any functionality or accessibility requirements. Hence, a clean-slate solution would require a wide range of largely unrelated functionalities beyond the core verifiability guarantees we aim to support. 

Second, election officials are often extremely resource-constrained~\cite{electionresources}. Practical overhead---administration, time, cost, and access to expertise---of adopting a completely new system would be prohibitive, and existing contracts and business relationships mean current systems are a sunk cost. As a result, adding to an existing system is much more feasible than replacing the entire voter registration database, and makes cryptographic verifiability accessible to a much wider audience of election officials.

Third, overhauling critical infrastructure like voting systems is subject to justifiable caution and takes time. Designs that are backward-compatible are more likely to be considered at all, deployed within a shorter time-frame, and eventually result in full scale adoption.
We note that this design this design pattern is common among election systems that have seen real-world adoption, e.g., ElectionGuard~\cite{electionguard}, which sits on top of existing casting-and-tallying infrastructure.}

\paragraph{2. Preserve baseline security.} Verifiability systems should preserve (or enhance) the security properties of the underlying voter registration system. That is, assuming that the underlying system satisfies some baseline level of completeness, soundness, and secrecy, adopting a cryptographic verifiability system should result in a decrease of these guarantees. For example, the new system should leak no additional information beyond that which is enforced by the access control of the base system, and it should allow for interoperation with maintenance and oversight entities with as much security as before its adoption. In particular, this requires \emph{fine-grained access control} individual voter records, as different entities are allowed access to different fields of voter data.

\paragraph{3. Enhance verifiability and verifiability.} \newdiff{The third main design goal is to meet the four verifiability guarantees outlined in Section~\ref{subsec:security-requirements}. A new system should increase confidence in the correctness, soundness, and privacy of the underlying voter registration system, by attesting cryptographically to stored data, updates, and maintenance procedures. Critically,} reaping the benefits of these verifiability properties should require as little technical expertise as possible (e.g., voters should not be required to remember cryptographic secrets).

\smallskip

These three goals span the four ``key challenges'' outlined in Section~\ref{sec:voter-reg-background}: the first goal addresses compatibility with existing analog and digital voter registration processes (challenges 1 and 2), the second goal ensures that \sysname respects existing policies 
and constraints on data protection and access control (challenge 4), and the third goal aims for strong verifiability irrespective of digital literacy (challenge 3).

\paragraph{Other requirements.} Besides the main goals described above, there are a few other requirements for \sysname. First, the registry should allow for \emph{efficient monitoring}: voters should not be required to verify their data at set times. (We assume that at least one party that a voter trusts verifies the correctness of every epoch.)
Second, we want to ensure that voters can easily verify their record without having to store \emph{any} secrets, and to ensure 
efficient key management for election officials.

\paragraph{Scope.}
Our design focuses exclusively on one aspect of voter registration systems, namely, \emph{data storage}. 
Other desirable goals that are part of distinct processes in the broader system are outside our scope, notably including real-world identity verification and verification of authenticity of voter data.
We assume that election officials, maintenance entities, and oversight entities have existing processes for such verification and treat these processes as black-box subroutines.

\section{\sysname: A Practical Prototype}\label{sec:vrlog}
\newdiff{We now describe \sysname, our design for a transparent voter registration database system. Aligned with the challenges for real-world adoption highlighted in Section~\ref{sec:voter-reg-background}, our goal with \sysname is to assemble the simplest set of cryptographic tools that are sufficient to meet our verifiability goals.}

\subsection{Background on Verifiable Registries}\label{subsec:technical-preliminaries}
\newcommand{\Setup}{\textsf{Setup}}
\newcommand{\Append}{\textsf{Append}}
\newcommand{\ProveL}{\textsf{ProveLookup}}
\newcommand{\ProveAO}{\textsf{ProveAppendOnly}}
\newcommand{\VerLookup}{\textsf{VerLookup}}
\newcommand{\VerAO}{\textsf{VerAppendOnly}}
\newcommand{\ProvePSI}{\textsf{ProvePSI}}
\newcommand{\VerPSI}{\textsf{VerPSI}}

A \textit{verifiable registry}~\cite{chase2019seemless,meiklejohn2020think,tomescu2019transparency} $\reg$ is a data structure maintained by a centralized server $S$ which stores (1) a directory of key-value pairs and (2) a history of updates for each key.\footnote{Some designs store a ``checkpointed'' history rather than a complete history (e.g.,~\cite{len2023optiks,malvai2023parakeet}); the details are beyond our present scope.} The registry is \emph{verifiable} in that the server can provide publicly verifiable cryptographic proofs of certain registry properties (primarily via \emph{inclusion}, \emph{non-inclusion}, and \emph{history} proofs).

Verifiable registries assume the existence of a public, append-only \emph{bulletin board} $\bulletin$, which is external to $\reg$, where $S$ periodically publishes a \emph{succinct commitment} to the registry. Such a bulletin board---which is a standard assumption for verifiable registries---can be implemented through a gossip protocol~\cite{syta2016keeping,meiklejohn2020think} or a public blockchain~\cite{tomescu2017catena}. Every commitment represents a new \emph{snapshot} of $\reg$, and we refer to the period of time between two snapshots as \emph{epochs}. $\bulletin$ serves as the central ``source of truth'' for the registry's state, hence providing its verifiability properties. \emph{Auditors}, which may be clients themselves, consistently monitor $\bulletin$, and use the published commitments to provably verify that each new version is a correct transition from the prior version, i.e., that the latter is a strict prefix of the former. In our setting, the registry's correctness only requires \emph{one} honest auditor per epoch.

Clients can then leverage $\bulletin$ to verify the correctness of the data in the registry. To do so, whenever they ask $S$ for data contained in $\reg$, they additionally request inclusion proofs and history proofs.
Auditors and clients have complementary roles assuring the correctness of the system: the former ensure that the high-level operation of the registry is honest, while the latter ensure that their individual data is correct.

\paragraph{API.} The main functions of a registry $\reg$ are as follows. For simplicity, we omit details not relevant to this paper.\footnote{See~\cite{chase2019seemless,meiklejohn2020think,malvai2023parakeet,len2023optiks} for a complete treatment.} 
\begin{newitemize}
    \item $\regupdate(\{k, v\})$: updates $\reg$ with the set of input pairs $\{k, v\}$, and creates a commitment $\regcom$ and a proof $\Pi^{upd}$ that attests that this update was performed correctly, then publishes $(\regcom,\Pi^{upd})$ to $\bulletin$. 
    \item $\reglookup(k)$: if $k\in\reg$, returns the latest value $v$ associated with key $k$ and an inclusion proof $\pi$; if $k\notin\reg$, returns a non-inclusion proof.
    \item $\regverlookup(\regcom, k, v, \pi)$: verifies the inclusion proof $\pi$ for $v$ under $k$, with respect to the snapshot of $\reg$ represented by $\regcom$.
    \item $\reghistory(k)$: if $k\in\reg$, returns all $n$ updates $[(v_i, t_i)]_{i \in [n]}$ to $k$ and the snapshot in which they occurred, alongside a history proof $\Pi^{hist}$.
    \item $\regverhistory(k, [(v_i, t_i)]_{i \in [n]}, \Pi^{hist}, [\regcom_i]_{i \in n})$: verifies the history proof $\Pi^{hist}$ for all $n$ updates to $k$, with respect to the snapshots of $\reg$ in which they occurred.    
    \item $\regaudit(\regcom_j, \regcom_{j+1}, \Pi^{upd})$: verifies the update proof $\Pi^{upd}$ with respect to two consecutive snapshots of $\reg$.
\end{newitemize}

\medskip

\paragraph{Security properties.} The security requirements of a registry 
consist of three core properties,
\iffullversion
explained informally below; see~\cite{chase2019seemless,meiklejohn2020think,malvai2023parakeet} for formal definitions.

\begin{newitemize}
\item \emph{Completeness}: if $\reg$ is updated honestly by $S$, then (1) for any key $k$ and any version of $\reg$, inclusion proofs (resp. history proofs) with respect to the correct latest value (resp. correct history of updates) associated with $k$ should successfully verify; and (2) auditing the transition from any version of the registry to the next one should successfully verify. 
\item \emph{Soundness}: assuming that there is at least one honest auditor in every epoch, querying the latest value associated with any key $k$ at any epoch $t$ should be consistent with a history proof for $k$ for any epoch ranges that include $t$. That is, a history proof will detect if $S$ ever reports an incorrect value for $k$. Note that a client is responsible for verifying the correctness of their historical data. This property guarantees that any other entities who fetch this client's data will receive an output consistent with the client's verification. 
\item \emph{Privacy}: the public commitments to $\reg$, update proofs, inclusion proofs, and history proofs, should reveal no additional information about the state of $\reg$. 
\end{newitemize}
\fi

\subsection{Design Overview and Notation}

Our design involves two main data structures maintained by election officials: the VRDB $\vrdb$, and a new verifiable registry $\reg$. We also introduce a public bulletin board $\bulletin$, which may or may not be managed by election officials.
\newdiff{In line with our goal of backward compatibility (Section~\ref{sec:voter-reg-background}), \sysname is an \emph{extension} to a jurisdiction's existing VRDB.}

We refer to the pre-existing voter registration system as the \emph{base system}, denoted by $\basesys$. The modules\footnote{As defined in Section~\ref{subsec:voter-reg-framework} under Voter Registration Modules.} of \sysname comprise \emph{extensions} of the respective modules of $\basesys$, as well as two additional modules that we introduce below. 

We denote the column labels of $\vrdb$ by $\vrdbcols = (\vrdbcol_1, ..., \vrdbcol_n)$ (which are jurisdiction-specific), the set of voters by $\voters = \{\voter_1, ..., \voter_m\}$, and voter $\voter_i$'s data in $\vrdb$ by a vector $\vrdb_{\voter_i} = (f_{i, 1}, ..., f_{i, n})$, where $f_{i, j}$ is $\voter_i$'s stored data under column~$\vrdbcol_j$. 

Next, we overview the extensions that our design \sysname makes to the base system $\basesys$. Details follow in Section~\ref{subsec:detailed-system}.

\paragraph{Setup and structure.}
\sysname associates to every $\voter$ a \emph{unique identifier}, which we denote by $\id_\voter$, and which leaks no information about the identity of the corresponding voter. These are assigned at the time a voter registers (or when \sysname is first incorporated into $\basesys$, for already registered voters). Each $\id_\voter$ is stored as a key of $\reg$, which maps to (obfuscated versions of) the corresponding voter's data $\vrdb_{\voter}$. Sensitive data is encrypted so that all registry content can be public; data deemed less sensitive may be stored in the clear to enhance transparency. We define the predicate $\public \colon \voter \times \vrdbcols \to \{0, 1\}$ as specifying whether a given field for a given voter should be encrypted. Our system is agnostic to which fields need to be protected, 
so is compatible with any jurisdictional policy regarding voter data protection.
We denote the most recent value associated with $\id$ in the registry as $\reg_\id$. 

\sysname then augments the base system's \Reg, \UpdateReg, \Maintenance, and \Oversight modules, by adding auxiliary steps involving $\reg$ to each module, as below.

\paragraph{Voter-initiated updates.} Recall that the last step in the \Reg and \UpdateReg modules is for election officials to update $\vrdb$ with the voter's new data. Our system enhances these modules with an additional step, whereby election officials add the update to a queue $\pool$ of all incoming changes of voter data (additions, deletions, and modifications). These updates are then pushed to $\reg$ in batches as part of a separate (periodic) process, which results in a new snapshot of the registry and the publication of latest commitment and update proof on a public bulletin board $\bulletin$. 
\iffullversion
    (The exact periodicity of pushes will vary by jurisdiction.)
\fi

\paragraph{Interoperability with third parties.} We denote the set of third parties (i.e., neither voters nor election infrastructure) that are approved to receive any voter data in this jurisdiction by $\thirdparties = \{\thirdparty_1, ..., \thirdparty_n\}$. 
This list generally includes the public, which has access to public voter data, and may include maintenance entities and oversight entities. 
Each entity in $\thirdparties$ is approved to receive a subset of fields from a subset of $\voters$, which we formalize via the predicate $\access \colon \thirdparties \times \voters \times \vrdbcols \to \{0, 1\}$, which outputs 1 if and only if the third party is allowed to access the column for the specified voter.\footnote{The reason this predicate takes the list of voters as input is that secrecy requirements may differ by voter, e.g. address confidentiality programs.}

In the base voter registration system, \basesys, election officials would directly send the voter data to  $\thirdparty$ that it is approved to access.
\sysname enhances this process as follows. For every $\voter$ that $\thirdparty$ is allowed to access at least one field of, election officials send (1) all of $\reg_{\id_\voter}$, alongside an \emph{inclusion proof} for it in $\reg$, and (2) some auxiliary, sensitive data that can be used to ``open'' the approved fields within $\reg_{\id_\voter}$. Then, $\thirdparty$ locally verifies the inclusion proof to ensure that $\reg_{\id_\voter}$ is consistent with the plaintext data, which can be checked by voters for correctness. Note that this check can occur post facto: voters need not have verified their data before $\thirdparty$ requests it; verifying a history proof, at some later point, is sufficient to detect any inconsistencies in prior epochs. Lastly, $\thirdparty$ opens the relevant fields of $\reg_{\id_\voter}$, and gains access to the plaintext data they need.

\paragraph{Data verification and system auditing.} \sysname introduces two new workflows to \basesys: \Query and \Audit. 
Through \Query, voters request inclusion proofs and history proofs for their data in $\reg$, which they can verify locally using the commitments posted in $\bulletin$ to ensure that their data is correct. Then, through \Audit, any entity can act as an \emph{auditor}, and verify the commitments and update proofs published by election officials to $\bulletin$, to ensure that $\reg$ has been updated correctly at every epoch in an append-only manner.

\subsection{Detailed System Description}\label{subsec:detailed-system}
\begin{figure}[t]
    \centering
    \includegraphics[width=\columnwidth]{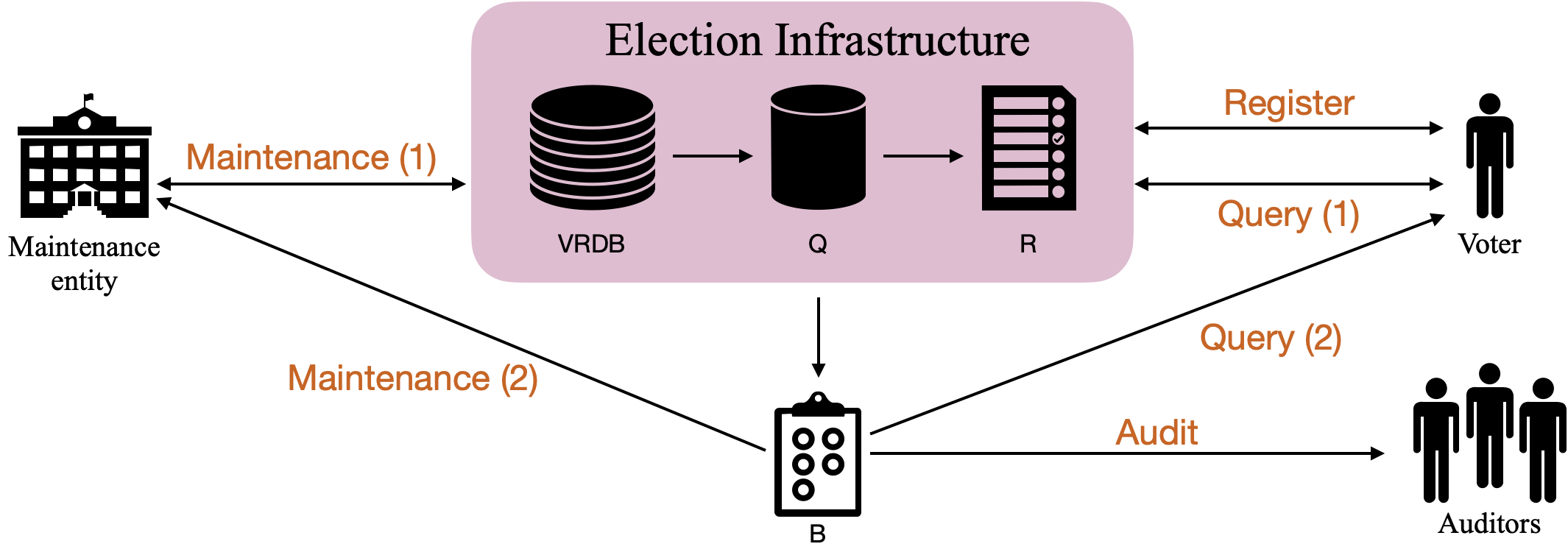}
    \caption{High-level summary of our architecture, where other third-parties besides maintenance entities are left implicit.
    } 
    \label{fig:system-fig}
\end{figure}

We now explain \sysname in detail. A graphical representation of our architecture is shown in Figure~\ref{fig:system-fig}.

\subsubsection{Voter-Initiated Updates}\label{subsubsec:registration}
\sysname enhances the base functionality of  $\basesys.\Reg$ in two key ways, which we explain below. 
\iffullversion
A formal description of this process can be found in Figure~\ref{fig:register-algo}.
\fi

First, upon successful verification of the prospective voter $\voter$'s eligibility, election officials use a secure pseudorandom function (PRF) $F$ to generate a random, anonymous identifier $\id_\voter$ for the voter, which reveals no information about their identity. To do so, they compute $\id_\voter = F(K_{id}, \vrdb_{\voter_{id}})$, where $K_{id}$ is a master PRF key held by election officials and $\voter_{id}$ is some existing identifier for the voter.\footnote{Most voter registration databases today have unique identifiers for voters. If not, one can generate unique identifiers prior to configuring \sysname.} 

After this, election officials compute an \emph{update record} for this transaction, which consists of the concatenation of the encrypted field values\footnote{Election officials should pad the plaintext fields to hide their lengths.}  within $\vrdb_\voter$ alongside additional metadata. The key used for each field is derived from the respective column label via a key-derivation function (KDF), using $\id_\voter$ and $\reg$'s current version number $e$ as a nonce. That is:
\begin{align*}
\centering
k_{j,e} &= \KDF(K_{kdf}, \id_\voter \concat \vrdbcol_j \concat e) \\
\reg_{\id_\voter} \coloneqq \updrec(\vrdb_{\voter}) &= \Enc_{k_{1,e}}(f_{1}) \concat ... \concat \Enc_{k_{n,e}}(f_{n}) \concat M,
\end{align*}
where $\Enc$ is a cryptographically-secure, \emph{key-committing}~\cite{franking} symmetric encryption algorithm, $\KDF$ is some key-derivation function (KDF) compatible with $\Enc$, $K_{kdf}$ is some master KDF key, and $M$ is additional metadata associated with this transaction. (Section~\ref{sec:security-analysis} discusses why the encryption needs to be key-committing.) Note that each individual voter has a different list of $n$ encryption keys, but election officials do \emph{not} need to store each one; they can be recomputed as needed. While it may appear this is a large number of keys, we note this is the lowest possible number of keys needed to both offer per-voter, per-field access control in addition to allowing key revocation. As keys do not need to be stored, but rather can be generated from one master key, this does not impede storage or computation (see Section~\ref{sec:impl} for our evaluation).

Voters need not store their own keys, satisfying our usability aims: voters can obtain their keys at any point by authenticating to the election official to retrieve their voter record.\footnote{While perhaps unusual for end-users to not need to store keys, in this case there is no added security benefit to voters storing their own keys. Election officials already have access to the full unencrypted voter record (as required by law), and voters can leverage the keys to verify their inclusion in the VRDB, so the management of keys by election officials does not diminish confidentiality nor verifiability of a voter's record, while enhancing usability.} The election official will provide the voter with their $n$ encryption keys, which the voter can use to verify their record. Further, note that the version number of $\reg$ serves as an implicit form of key rotation: subsequent updates to $\voter$'s data will use fresh keys as a result of the changing nonce in $\KDF$.

The metadata $M$ contains four pieces: (1) a timestamp indicating the time of registration and the version number of $\reg$, (2) an $\add$ operation code indicating the operation being performed on this voter's data, a (3) a \emph{group signature} signed with a key linked to the employee or machine who handled this registration (for traceability purposes), and (4) additional jurisdiction-specific data as needed.

Finally, the modified \Reg workflow concludes with election officials adding a tuple $(\id_\voter, \reg_{\id_\voter})$ to a \emph{queue of updates} $\pool$, which collects all new registrations, as well as modifications and deletions of existing registrations.

The modified \sysname.\UpdateReg workflow is the same as for registering, except that the $\update$ or $\delete$ operation code is used instead of $\add$. In particular, $\updrec$ is computed the same way, and all fields are re-encrypted (for reasons we explain in Section~\ref{sec:security-analysis}). This module also concludes with a new update record added to $\pool$.

\iffullversion
\renewcommand{\algorithmicrequire}{\textbf{Input:}}
\begin{figure}[t]
\begin{algorithm}[H]
\caption{\sysname.\Reg}
\begin{algorithmic}
\small
\Require $\vrdb_{\voter} \coloneqq (f_{1}, ..., f_{n}), M, e$
\State run $\basesys.\Reg(\vrdb_{\voter})$ \Comment{Abort if base system aborts.}
\State $\id_\voter \gets F(K_{id}, \vrdb_{\voter})$
\State $r \gets \bot$
\For{$j \gets$ 1 to $n$}
\If{$\public(\voter, \vrdbcol_j) = 1$}
\State $r \gets r \concat f_j$
\Else \State $k_{j,e} \gets \KDF(K_{kdf}, \id_\voter \concat \vrdbcol_j \concat e)$ 
\State $r \gets r \concat \Enc_{k_{j,e}}(f_{j})$
\EndIf
\EndFor
\State $M \gets M \concat \add$
\State $\pool$.add($\id_\voter, r$)
\State \Return $\id_\voter$
\end{algorithmic}
\end{algorithm}
\caption{The main steps involved in registering new voters using \sysname.}\label{fig:register-algo}
\vspace{-0.4cm}
\end{figure}
\else
\fi

\paragraph{Updating the verifiable registry.} The queue of update records $\pool$ captures all new registrations, voter-initiated updates, and modifications that arise via routine list maintenance activities (we discuss these in the subsequent section). Periodically, as specified by each jurisdiction, the contents of $\pool$ are pushed to $\reg$, which completes the lifecycle of an update to voter data. Note that $\vrdb$ and $\reg$ are not always consistent, as the former contains updates that are still in $\pool$. An update is not considered finalized until it reaches $\reg$.\footnote{In practice $\reg$ can be updated very frequently, e.g., daily or hourly.}

To process $\pool$, election officials first update $\reg$ with $\pool$ itself as input by calling $(\regcom, \Pi^{upd}) \leftarrow \regupdate(\pool)$. $\pool$ can then be supplied to $\regupdate$, as it consists of a series of key-value pairs $(\id_i, \reg_\id)$. Then, election officials cryptographically sign and publish $(\regcom, \Pi^{upd})$ to the bulletin board $\bulletin$, completing the transition to the next epoch.

\subsubsection{Interoperability with Third Parties}\label{subsubsec:list-maintenance}
Next, we describe how third parties (such as maintenance or oversight entities) can interface with \sysname to receive subsets of voters' data. Interfacing with a third party $\thirdparty$ involves three main steps, in the base system: (1) election officials send a (plaintext) subset of the voter records to 
\iffullversion
    $\thirdparty$ (e.g., as determined by the jurisdiction's data access control policy as defined in~\cite{cable2023systematization});
\fi
(2) $\thirdparty$ processes this data locally, and reports back to either the election official (for maintenance entities) or the public (for oversight entities); and (3), in the case of maintenance entities, election officials processes the response and update the VRDB if needed.

We enhance the existing process as follows, 
\iffullversion
which is formalized in Figure~\ref{fig:maintenance-algo}. 
\fi
First, in step (1), election officials send the following to $\thirdparty$: a (cryptographically-signed) inclusion proof for $\reg_{\id_{\voter_i}}$, for every $\voter \in \voters$ for whom $\thirdparty$ is approved to access to at least one field; and the decryption keys $k_{i, j}$ for every specific, non-public field of each voter that $\thirdparty$ is approved to access.
Then, in step (2), $\thirdparty$ locally verifies that the inclusion proof $\pi_i$ for every $\reg_{\id_{\voter_i}}$ is valid with respect to the latest published commitment in $\bulletin$. If these checks fail, $\thirdparty$ aborts the protocol and makes a notice through an out-of-band reporting mechanism (see Section~\ref{subsubsec:data-verification-system-audit}). Otherwise, for each $k_{i, j}$, $\thirdparty$ decrypts the $j$-th ciphertext contained within $\reg_{\id_{\voter_i}}$. Thus, at the end of this process, $\thirdparty$ acquires the same plaintext fields it would have received in the base protocol, but with additional guarantees that the data they received is consistent with the contents of VRDB.

Lastly, after step (3), if an update is made, election officials craft an update record for this transaction. This update record gets collected in the pool of updates $\pool$, waiting to be pushed to $\reg$ in the transition to the next epoch. The granularity of update records can be a transparency parameter specified by jurisdictions. Using the epoch number as part of the encryption key derivation process --- our implicit form of key rotation --- allows us to support changes in access control that may arise: if $\thirdparty$ is no longer allowed to access field $f_{i, j}$, its old decryption keys will no longer work for subsequent updates.

\iffullversion
\newcommand{\vt}{\mathsf{voters}}
\newcommand{\p}{\mathsf{proofs}}
\newcommand{\keys}{\mathsf{keys}}
\newcommand{\f}{\mathsf{fields}}

\renewcommand{\algorithmicrequire}{\textbf{Input:}}
\begin{figure}[t]
\begin{algorithm}[H]
\caption{\sysname.\Maintenance}
\begin{algorithmic}
\small
    \Require $\thirdparty, \vrdb, \reg, \voters, \vrdbcols$
    \State run step (1) of \basesys.\Maintenance \Comment{Abort if base module aborts.}\small
    \State $\vt, \p, \keys \gets (), (), (), ()$
    \For{$i \gets 1$ to $m$}
    \State approved$\gets \false$
    \For{$j \gets 1$ to $n$}
    \If{$\access(\thirdparty, \voter_i, \vrdbcol_j) = 1 \wedge \public(\voter_i, \vrdbcol_j) = 0$}
    \State approved$\gets \true$
    \State \Comment{$e$ is retrieved from $\reg_{\id_{\voter_i}}$}
    \State $\keys \gets \keys \cup \big(\KDF(k_{kdf}, \vrdbcol_j \concat \id_{\voter_i} \concat e)\big)$
    \Else \State $\keys \gets (\bot)$
    \EndIf
    \EndFor
    \If{approved}
    \State $\vt \gets \vt \cup (\id_{\voter_i})$
    \State $\p \gets \p \cup \big(\reglookup(\id_{\voter_i})\big)$
    \Else \State $\vt, \p \gets \vt \cup (\bot), \p \cup (\bot)$
    \EndIf
    \EndFor
    \State \Return $(\vt, \p, \keys)$
\end{algorithmic}
\medskip
\hrule
\begin{algorithmic}
\small
    \Require $\vrdbcols, \regcom, (\vt, \p, \keys)$
    \State $\f \gets ()$
    \For{$i \gets 1$ to len$(\vt)$}
    \If{$\vt_i \neq \bot \wedge \regverlookup(\regcom, \vt_i, \p_i) = 0$}
    \State \textbf{abort}
    \EndIf
    \For{$j \gets 1$  to $n$}
    \State $ix \gets i \times n + j$
    \If{$\keys_{ix} \neq \bot\ $}
    \State $\f \gets \f \cup (\Dec_{\keys_{ix}}(\p_{i}[0][j]))$
    \EndIf 
    \EndFor
    \EndFor
    \State run step (2) of \basesys.\Maintenance \Comment{Abort if base module aborts.}
    \State \Return $\f$
\end{algorithmic}
\end{algorithm}
\caption{The main steps involved in interfacing with third parties using \sysname, as run by election officials (top) and third-parties (bottom). Here we show the \Maintenance module as an example, but the steps are general for any interaction.}\label{fig:maintenance-algo}
\vspace{-0.4cm}
\end{figure}

\else
\fi

\subsubsection{Data Verification and System Audit}\label{subsubsec:data-verification-system-audit}
As discussed in Section~\ref{subsec:technical-preliminaries}, the transparency properties of a verifiable registry rely on two key assumptions: (1) that data owners are able to verify the correctness of their data, and (2) at least one honest auditor per epoch verifying the published commitments, to ensure that updates were performed correctly.\footnote{In this case of elections, this is a reasonable assumption, as existing groups (e.g., nonprofits, advocacy groups, etc.) who often already assess databases can act as auditors.} Our system supports these two operations via the simple \sysname.\Query and \sysname.\Audit workflows.

In the \sysname.\Query module, $\voter$ first specifies a range of time for which they want to verify their data; this could be the latest epoch, the entire history of their data, and anything in between.  In response, election officials send to the voter their (1) their identifier $\id$; (2) their history of all $e$ data updates, and a proof attesting to this history, computed as
\[\Big(\big[(\reg_{\id_i}, t_i)\big]_{i \in [e]}, \Pi^{hist}\Big) \leftarrow \reghistory(\id);\]
and (3) all $n*e$ secret keys $\{k_{j,i}\}_{j \in [n], i \in [e]}$ representing every included epoch, which are easily re-computed.
The proof $\Pi^{hist}$ is additionally signed by the election official with a digital signature, in order for voters not to be able to incorrectly claim that the election official supplied incorrect proofs.

Upon receiving this data, the voter verifies correctness by running the following subroutine (shown in 
\iffullversion
Figure~\ref{fig:query-algo}). 
\fi
First, they verify that the history proof is correct with respect to the commitments in $\bulletin$ for the update epochs. If that check passes, the voter proceeds to verify the correctness of each independent field of each data update: for all $e$ versions of $\reg_{\id}$, decrypt each individual field with corresponding key $k_{i,j}$, and verify the resulting plaintext field to ensure that there are no unexpected updates.

\iffullversion
\renewcommand{\algorithmicrequire}{\textbf{Input:}}
\begin{figure}[t]
\begin{algorithm}[H]
\caption{\sysname.\Query}
\begin{algorithmic}
\small
\Require $\id, \{k_{j,i}\}_{j \in [n], i \in [e]}, \big[(\reg_{\id_i}, t_i)\big]_{i \in [e]}, \Pi^{hist}, [\regcom_i]_{i \in [e]}$
\If{$\regverhistory(\id, \big[(\reg_{\id_i}, t_i)\big]_{i \in [e]}, \Pi^{hist}, [\regcom_i]_{i \in [e]}) = 0$}
\State \textbf{abort}
\EndIf
\For{$i \gets 1$ to $e$}
\For{$j \gets 1$ to $n$}
\If{$\public(\voter, \vrdbcol) = 1$} $f_j = \reg_{\id_i}[j]$
\Else\ $f_{j} = \Dec_{k_{j,i}}(\reg_{\id_i}[j])$
\State \Comment{Abort if $f_j$ is incorrect.}
\EndIf
\EndFor
\EndFor
\State \Return
\end{algorithmic}
\end{algorithm}
\caption{Additional \Query module introduced by \sysname.}\label{fig:query-algo}
\vspace{-0.4cm}
\end{figure}
\else
\fi

In the \sysname.\Audit module, auditors select a pair of consecutively-published commitments $(\regcom_i, \Pi^{upd}_i)$ and $(\regcom_{i+1}, \Pi^{upd}_{i+1})$ in $\bulletin$, and verify that this transition was performed correctly by computing $\regaudit(\regcom_i, \regcom_{i+1}, \Pi^{upd}_{i+1})$.

\paragraph{Dispute resolution.} We assume an out-of-band reporting mechanism, by which entities in the system can report that some data in (or activity on) $\reg$ should be corrected. There are three types of disputes that can arise in our system: (1) voters who detect modifications to their data when verifying history proofs, as part of the \sysname.\Query; (2) maintenance and oversight entities who detect that they received incorrect data from election officials, as part of the \sysname.\Maintenance and \sysname.\Oversight; and (3) auditors who detect that $\reg$ was not updated correctly when verifying update proofs as part of the \sysname.\Audit. In any of these cases, $\bulletin$ can be used to resolve any disputes, as any proofs that cannot be verified confirm that $\reg$ was not updated correctly. Commitments and proofs are signed, protecting against false claims of malfeasance of election officials by producing a fake proof.

\section{Security Analysis}\label{sec:security-analysis}
We now formalize the security requirements for cryptographic registration transparency, and prove that \sysname meets these definitions. Table~\ref{fig:entities-assumptions} summarizes the security guarantees of our system and the parties and assumptions relevant to each.

Recall from Section~\ref{subsec:design-goals}
that we aim to support two main security goals: \sysname should (1) preserve the base system's completeness, soundness, and secrecy properties; and (2) increase the transparency of the system. We address each below. 

\begin{table*}
    \centering
    \footnotesize
    \renewcommand{\arraystretch}{1.1}
    \begin{tabularx}{\textwidth} { 
  >{\raggedright\arraybackslash}p{0.15\textwidth} 
  >{\raggedright\arraybackslash}X 
  >{\raggedright\arraybackslash}X
  >{\raggedright\arraybackslash}X
  >{\raggedright\arraybackslash}X}
     \toprule
      \textbf{Entity} & \textbf{Responsibilities} & \textbf{Mechanisms} & \textbf{Additional Assumptions} & \textbf{Protects Against}\\
     \midrule
     Auditors & Ensure that $\reg$ is updated correctly. & Verify the commitments and update proofs  published to $\bulletin$. & There is at least one honest auditor per epoch. & Split-view attacks from election officials. \\
     Voters &  Ensure the correctness of their personal data in $\reg$. & Request and verify history proofs for their data. & Every epoch is contained in at least one history proof. & Maintenance and oversight entities receiving incorrect data from election officials. \\
     Maintenance entities & Verify the consistency of the data received from election officials. & Verify inclusion proofs with respect to the commitments in $\bulletin$. & \\ 
     \bottomrule
\end{tabularx}
\caption{Summary of assumptions and responsibilities for each entity in each modified module, and resulting guarantees.}
\vspace{-0.3cm}
\label{fig:entities-assumptions}
\end{table*}

\subsection{Preserving Properties of the Base System}

To show that our system does not diminish the secrecy of the underlying voter registration system, we must establish that the parties involved in our protocols do not learn any more information than they would have learned in the base modules. Our analysis examines three key contexts and shows that \sysname does not leak information in all cases: 
(1) registered voters participating in the \Query module, (2) maintenance and oversight entities who participate in the \Maintenance and \Oversight modules, and (3) auditors (or any member of the public) who inspect the bulletin board via the \Audit module.

\paragraph{Informal theorems.}
Theorems~\ref{thm:voter-secrecy} and \ref{thm:public-secrecy} represent the following guarantees respectively: (1) for any voter $\voter \in \voters$, the \Query module does not reveal any voter's data from $\vrdb$ beyond $\vrdb_{\voter}$, and (2) for any member of the public, $\reg$ and $\bulletin$ do not reveal any fields $f_{i, j} \in \vrdb$ for which $\public(\voter_i, \vrdbcol_i) = 0$. 
A third theorem very similar to Theorem~\ref{thm:voter-secrecy}, but focused on third parties, is as follows: (3)
for any third-party $\thirdparty \in \thirdparties$, their interaction with $\reg$ does not reveal any fields $f_{i, j} \in \vrdb$ for which $\access(\thirdparty, \voter_i, \vrdbcol_j) = 0$.
Below, we formalize Theorems~\ref{thm:voter-secrecy} and \ref{thm:public-secrecy}.
We omit the third theorem as both its statement and proof are closely analogous to Theorem~\ref{thm:voter-secrecy}.

\paragraph{Assumptions.}
All theorems below assume: (1) the registry $\reg$ satisfies the standard \emph{completeness}, \emph{soundness}, and \emph{privacy} properties of verifiable registries (Section~\ref{subsec:technical-preliminaries}); (2) all entities satisfy our assumptions (Table~\ref{fig:entities-assumptions}); (3) the cryptographic primitives in our protocols satisfy 
standard security definitions; and (4)  the encryption scheme used in $\reg$ is key-committing\cite{franking}.

We write $\sysname.\Modules$ to denote the collection of all modules in $\sysname$.
\vspace{-0.05em}

\begin{thm}[Voter Secrecy]\label{thm:voter-secrecy}
    Let $\vrdbcols$ be any arbitrary column labels, and let $\{p_\vrdbcol\}_{\vrdbcol \in \vrdbcols}$ be any set of probability density functions for their distribution of possible values. Then, for any Turing Machines $O_1, O_2$ and any adversary $\advA$, there exists a negligible function $\epsilon$ such that, for any $k \in \mathbb{N}$:
    \begin{equation*}
    \small
        Pr\left[   
\begin{array}{llll}
\voters, \vrdb, \reg \gets \emptyset \\
O_1(\voters, \vrdb, \reg)^{\sysname.\Modules(\cdot)} \\
\advA(\reg, \voters) = \voter \in \voters \\
O_2(\voters, \vrdb, \reg)^{\sysname.\Modules(\cdot)} \\
\exists\ f_{i, j} \in \vrdb : \voter_i \neq \voter \wedge \\ \big|p_{\vrdbcol_j}(f_{i, j}) - \advA^{\sysname.\Query(\voter)}(\reg) = f_{i, j}\big|
\end{array}
\right] < \epsilon(k)\ .
    \end{equation*}
    
\end{thm}
\begin{pfsketch}
As part of the \Query module, $\voter$ receives their identifier $\id$, their data $\reg_\id$ stored in the registry, the secret keys used to to decrypt this data, and a history proof $\Pi^{hist}$. Since $\id$ is computed only as a function of $\voter_{id}$, and $F$ is a secure PRF, this identifier does not leak any information about the data of other voters. Then, the security of $\KDF$ guarantees that knowledge of $\voter$'s decryption keys does not reveal information about the decryption keys of other voters, as these were computed using different seeds. Lastly, the \emph{privacy} properties of $\reg$ guarantee that $\reg_\id$ and $\Pi^{hist}$ do not leak any information about the state of the other values contained in $\reg$.
\end{pfsketch}

\begin{thm}[Public Secrecy]
\label{thm:public-secrecy}
    Let $\vrdbcols$ be arbitrary column labels. For any Turing Machines $O_1, O_2$ and any adversary $\advA$, there exists a negligible function $\epsilon$ such that, for any $k \in \mathbb{N}$:
    \begin{equation*}
    \small
        Pr\left[   
\begin{array}{llll}
\voters_0, \vrdb_0, \reg_0, \bulletin_0, \voters_1, \vrdb_1, \reg_1, \bulletin_1 \gets \emptyset \\
O_1(\voters_0, \vrdb_0, \reg_0)^{\sysname.\Modules(\cdot)} \\
O_2(\voters_1, \vrdb_1, \reg_1)^{\sysname.\Modules(\cdot)} \\
b \assignRand \{0, 1\} \\
\advA(\reg_b, \bulletin_b) = b \wedge \reg_0 \equiv \reg_1
\end{array}
\right] < \epsilon(k)\ ,
    \end{equation*}
where $\reg_0 \equiv \reg_1$ denotes that $\reg_0,\reg_1$ are indistinguishable based on public information (i.e., public voter data and public metadata about registry activity).\footnote{If $\reg_0 \not\equiv \reg_1$, the two sequences of updates are trivially distinguishable.}
\end{thm}

The proof sketch for Theorem~\ref{thm:public-secrecy} is in Appendix~\ref{a:more-thms}.

\subsection{Increasing Transparency}
Our construction enhances the transparency of the voter registration system in three ways: (1) voters detecting if their data has been modified or deleted without their knowledge, (2) third-parties detecting if the data they received from election officials is not consistent with the VRDB, and (3) the public at large verifying the consistency of the registry. 

\paragraph{Informal theorems.}
Theorem~\ref{thm:voter-transparency} represents the following guarantee: (1) for any voter $\voter \in \voters$, election officials are not able to produce two proofs for two different values of $\voter$'s data in a way that will not be detectable by $\voter$.
Our two additional transparency theorems, stated informally, are: (2) any member of the public can ensure that no entries have been removed or modified from $\reg$; and (3) for any third-party $\thirdparty \in \thirdparties$ and voter $\voter$, election officials are not able to reveal a field of $\vrdb$ to $\thirdparty$ in a way that is not consistent with the data verified by $\voter$.

\begin{thm}[Voter Transparency]
\label{thm:voter-transparency}
Let $\vrdbcols$ be any arbitrary column labels. Then, for any Turing Machines $O_1, O_2$ and any adversary $\advA$, there exists a negligible function $\epsilon$ such that, for any $k \in \mathbb{N}$:
    \begin{equation*}
    \footnotesize
        Pr\left[   
\begin{array}{llll}
\voters_0, \vrdb_0, \reg_0, \bulletin_0, \voters_1, \vrdb_1, \reg_1, \bulletin_1 \gets \emptyset \\
O_1(\voters_0, \vrdb_0, \reg_0)^{\sysname.\Modules(\cdot)} \\
O_2(\voters_1, \vrdb_1, \reg_1)^{\sysname.\Modules(\cdot)} \\
b \assignRand \{0, 1\} \\
\advA(\vrdb, \reg, \voters) = \pi_0, \pi_1 \\
\exists\ \voter \in \voters, [\regcom_i]_{i \in E} \subseteq \bulletin_b : \reg_{0, \id_\voter} \neq \reg_{1, \id_\voter} \wedge \\ 

\scriptsize
\sysname.\Query(\id_\voter{,} \big[(\reg_{0, \id_\voter}{,} t_i)\big]_{i \in E}{,} \pi_0{,} [\regcom_i]_{i \in E}) {\neq} \bot \wedge \\

\sysname.\Query (\id_\voter{,} \big[(\reg_{1, \id_\voter}{,} t_i)\big]_{i \in E}{,} \pi_1{,} [\regcom_i]_{i \in E}) {\neq} \bot
\end{array}
\right] {<} \epsilon(k)\ .
    \end{equation*}
\end{thm}

See Appendix~\ref{a:more-thms} for the proof sketch for Theorem~\ref{thm:voter-transparency} and further discussion of the other two theorems.
\section{\sysnamepriv: Privacy-Enhancing Extension}\label{sec:privacy-ehancing-extensions}

So far, we have seen that \sysname \emph{preserves} the secrecy guarantees of the underlying voter registration system. In this section, 
we describe an enhancement that \emph{increases} privacy guarantees relative to the underlying voter registration system.

\paragraph{Deduplication in theory.} A key task within the \Maintenance module consists of jurisdictions comparing their voter records in order to detect duplicate voters (as also discussed in Section~\ref{sec:voter-reg-background}). That is, jurisdictions $A$ and $B$ have voter lists $L_A$ and $L_B$, and want to compute the intersection $L_A \cap L_B$. 

So far, this sounds like a classic use case for \emph{private set intersection} (PSI): a class of cryptographic protocols that allow $A$ and $B$ to securely compute the intersection $L_A\cap L_B$ while provably revealing no additional information, and without relying on a trusted party.

\paragraph{Deduplication in practice.}
Many U.S. states\footnote{As of Feb. 2024, ERIC has 24 states and D.C. as members~\cite{ericfaq}.} participate in deduplication mediated by the non-profit ERIC, which acts in the role of a trusted third party (as noted earlier in Sections~\ref{sec:related-work} and \ref{sec:voter-reg-background}). ERIC works as follows \cite{ERIC-pdf}: (1) participating states periodically submit their voter records 
to ERIC; (2) ERIC compares participating states' data with each other, official death data from the Social Security Administration, and official change-of-address data from the USPS; and  (3) ERIC notifies states of records that are possible duplicates with another state or otherwise appear to be outdated. States submit their data in plaintext, except for three fields that ERIC considers sensitive,\footnote{ERIC states that driver's license/state ID numbers, social security numbers, and birth dates are considered sensitive.} to which states apply a keyed hash before transmission, using a hardware security module that ERIC provides. The comparison step is performed by ERIC using commercial record-linkage software from Senzing.\footnote{It appears that Senzing does not have access to the voter records \cite{Senzing-API-site,ERIC-pennsylvania}.}

ERIC performs a valuable function, and represents an important collaboration between member states. \sysname is compatible with systems like ERIC: integrating with ERIC would require no extra work from ERIC or other states, while still providing an additional check that data ERIC handles is the same as recorded in that state's verifiable registry.

\paragraph{Improvements.}
ERIC's approach involves the transfer of more voter data than necessary using other cryptographic techniques.
Our proposal \sysnamepriv enhances \sysname to support \emph{direct input verification for cross-database secure computation} for deduplication as well as more complex cross-database checks, strengthening confidentiality \emph{and} verifiability.

\iffullversion
    In practice, PSI is not up to our task. PSI protocols are designed to find exactly equal records, whereas detecting voter records that are \emph{similar enough} to be likely duplicates is essential in our setting. For example, common issues include misspellings, abbreviations, name changes, and other outdated information. The similarity measure is complex: some of these do not correspond to a mathematically intuitive ``closeness'' metric (e.g., a name change). Moreover, it is critical that our protocols effectively compare records between databases of different formats that may contain different fields.
\fi

As such, we are facing the more general problem of \emph{privacy-preserving record linkage} (PPRL), which allows two parties to detect possible duplicate records --- for which the information in each party's database may look different --- without revealing any information about non-likely-duplicates.\footnote{PSI is a special case of PPRL (i.e., when the data for each entity is exactly the same in both databases), though the PSI and PPRL literature seem to have surprisingly few explicit connections.} These guarantees hold in an honest-but-curious security model or, for some protocols, even malicious security~\cite{christophides2020overview}.

\subsection{\sysnamepriv Framework}\label{subsec:privacy-extension-framework}
Jurisdictions could, of course, run a PPRL protocol without using \sysname at all. The advantage of connecting PPRL with \sysname is \emph{verifiable inputs}. Standard secure computation protocols (including PPRL) do not guarantee input correctness; instead, all security guarantees are \emph{relative} to the (secret) inputs that parties bring to the protocol~\cite{lindell2020secure,hall2010privacy}. 

\paragraph{PPRL.}
A PPRL protocol breaks down into two main steps.

\begin{newenum}
    \item \emph{Encoding}: Each entity pre-processes their input data, creating an encoding\footnote{Examples of encoding algorithms include Bloom filters and encryption.} that reveals nothing about the plaintext data, but is still suitable for approximate matching in the next step. We denote this step by $\encode(L)$.
    \item \emph{Matching}: The entities execute a protocol on the encoded data that results in the detection of elements that are likely to represent the same entity. We denote this step by $\match(\encode(L_a), \encode(L_B))$.
\end{newenum}
\iffullversion
    After the matching step, there is generally a manual comparison to check which of the output records actually correspond to the same entities. 
    We leave this step implicit hereafter. We define a PPRL protocol $P$ to be this pair of algorithms: $P \coloneqq (\encode, \match)$.\footnote{This notation, while sufficient for our purposes, glosses over many details. See surveys such as~\cite{gkoulalas2021modern,hall2010privacy,vatsalan2013taxonomy,vatsalan2017privacy} for a more comprehensive treatment.}
\fi

\paragraph{Enhancements to \sysname.} The key idea of \sysnamepriv is for jurisdictions to run a PPRL protocol $P \coloneqq (\encode, \match)$ at the start of the \sysname.\Maintenance module, but such that their encoded input data is \emph{stored directly in their registry}, which is verified by voters as part of the \sysname.\Query module. In more detail, \sysnamepriv makes three modifications to \sysname. First, the format of update records is modified such that $\encode(f_{i, j})$ is appended to $\Enc_{k_{i, j}}(f_{i, j})$ for each (non-public) field in the VRDB. Then, in the \sysname.\Maintenance module, whenever the two entities involved are two jurisdictions $A$ and $B$, they engage in some PPRL protocol $P$, such that \emph{$A$ retrieves $B$'s encoded list directly from $B$'s registry} (and vice-versa), instead of receiving it from $B$ as its input to $P$. Lastly, in the \sysname.\Query module, after the voter decrypts each received ciphertext and verifies the correctness of their data, they additionally verify\footnote{The details of this verification depend on the exact encoding scheme being used, which may require auxiliary data from election officials.} that the encoded data is consistent with the decrypted data. We show the full modifications that \sysnamepriv makes to the modules of \sysname in Appendix~\ref{a:sysnamepriv-modules}.

\paragraph{Using \sysnamepriv with ERIC.} 
\sysnamepriv remains compatible with ERIC.
Using \sysname with ERIC already provides assurance that the \emph{plaintext} voter data processed by ERIC is the same as in states' voter lists, but these guarantees do not extend to voter data that is not public (and thus not stored in plaintext in \sysname). This gap is filled by \sysnamepriv, which when used with ERIC provides transparency for both public (plaintext) voter data and non-public voter data.

\paragraph{Performance.}
The performance of PPRL protocols is generally judged along three axes: \emph{privacy}, \emph{scalability}, and \emph{linkage quality}.
\sysnamepriv makes black-box use of the input PPRL protocol $P$, and inherits its performance on all three axes directly from $P$. 
Thus, as PPRL protocols improve in privacy, scalability, and quality, \sysnamepriv will improve alongside.

There has been significant work evaluating the performance of PPRL protocols: see, e.g.,~\cite{mirel2022methodological,randall2022blinded,pprlsperformance} (specific analyses) and~\cite{gkoulalas2021modern,hall2010privacy,vatsalan2013taxonomy,vatsalan2017privacy} (surveys). Recent evaluations show commercial PPRL tools 
can achieve a linkage quality of upwards of 90\% on precision, 85\% on recall, 90\% on F-score, and 90\% on accuracy, even with data of only ``moderate quality and completeness,'' while processing millions of records in less than a day using a single core~\cite{pprlsperformance}. 
Large-scale PPRL experiments have been deployed in other contexts, such as deduplicating $\sim$170m patient health records~\cite{marsolo2023assessing}.
As list maintenance operations occur infrequently, protocol execution in a matter of hours is acceptable in our setting. 

Finally, on a fourth metric, \emph{storage}: \sysnamepriv's privacy enhancement comes at the cost of additional overhead for storing one additional encoding per field in the log, which approximately doubles the storage usage of the log.

\iffullversion
    \paragraph{Generalizing to secure computation.} The two-step protocol structure we have described --- that is, a local encoding step on parties' input data followed by some joint computation over the encoded data --- is a high-level structure common to many secure multi-party computation (MPC) protocols, including ones that support arbitrary computations.\footnote{This is not true for \emph{all} MPC protocols (e.g., in garbled circuits, the encoding step depends on the function and input data to be securely computed).} 
    \sysnamepriv can work with any MPC protocol that has this structure: parties store encoded inputs directly on the registry, which is verified by voters for correctness, and fetched by other parties at the start of the MPC protocol. Thus, \sysnamepriv can support input validation for secure computation protocols beyond PSI or PPRL, creating the potential for more complex cross-database computations involving verified voter data. \newdiff{Exploring this idea more generally, i.e., the use of verifiable registries to validate inputs to an MPC protocol, is an interesting direction for future work.}
\fi

\section{Implementation}
\label{sec:impl}
To show that cryptographic registration transparency is practical, we implemented and evaluated a prototype of \sysname, and demonstrate that it is capable of scaling at low cost to databases the size of the largest U.S. state.

\label{times}
\begin{figure}[t]
    \centering
    \includegraphics[width=8cm]{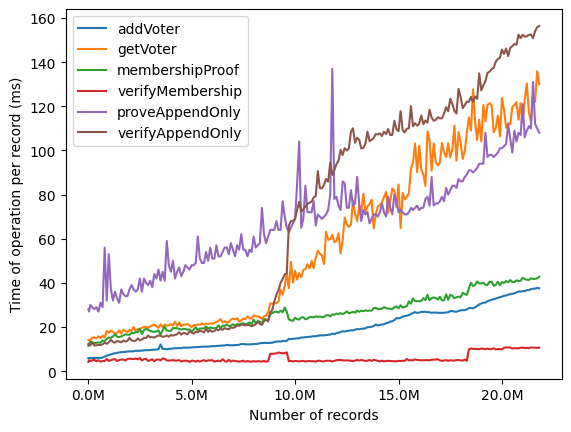}

    \caption{\centering Time per operation given $n$ 1KB records.}
    \label{fig:times}
\end{figure}

\paragraph{Prototype.}
We implement \sysname as a verifiable log-backed map using the Trillian library~\cite{trillian}.\footnote{In 2021, Trillian removed support for verifiable maps due to performance issues. We use the latest version prior to their removal (1.3.12).} In particular, we construct a verifiable log-backed map using a verifiable log to store mutations, a verifiable map which includes those mutations, and another verifiable log to store a history of signed root hashes of the map. Using the description of \sysname in Section \ref{subsec:detailed-system}, the verifiable log implements $\vrdb$ and the verifiable log-backed map implements $\reg$, with the publicly-accessible property of the verifiable log-backed map representing $\bulletin$. Our \href{https://github.com/cablej/vrlog}{open source implementation} consists of over 1,000 lines of code, written in Go, and fully implements the \sysname protocols.

The code includes a server which exposes an HTTP API that allows retrieving, adding and updating voters to satisfy the above properties, in addition to inclusion proofs of voter data, and proofs that the log is append only. Separately, the Trillian log and map servers are exposed in a read-only mode, which allows any party to access data and obtain proofs and verify data in the verifiable log-backed map. The API for write operations is intended to be accessible only directly by the election official. In practice, we expect that an election official would make calls to the API from the backend of their existing voter registration database, e.g., to add a voter to the map after they are added to the database.

We implement a key-committing encryption function for $\Enc$ by appending a hash of the key to the ciphertext. This is a standard transformation (``Encrypt-then-HMAC'') and only marginally increases storage overhead~\cite{franking}. We also provide an option for election officials to specify a fixed size for each field, which will then pad the stored data so as to resist leaking the length of the field (e.g., the length of a voter's name).

\paragraph{Evaluation.}
At a relatively low cost, \sysname is able to support databases that exceed the size of the largest states in the U.S.\footnote{California has 22 million registered voters~\cite{CARegistration}. We did not go beyond this, though it is clear the system can support hundreds of millions of voters.}
We operate on an AWS EC2 \texttt{t2.large} instance and AWS RDS Aurora MySQL \texttt{db.t4g.large} server. Excluding storage costs (which are minimal), running \sysname continually for a month on this infrastructure costs roughly \$160. 

We measure the speed of each operation and required total storage. 
Figure~\ref{fig:times} shows the average time to complete each operation given the number of entries in the log (i.e., adding, updating, proving/verifying voters, and proving/verifying that the log is append-only). The `proveAppendOnly' operation is for the entire database, i.e., it takes approximately 80 ms to prove the entire database is append only with 15M records, while other operations are per record. These timings scale linearly with added resources; we expect that a larger jurisdiction could see significant gains with minimal further investment.

The storage required for the log grows linearly with every operation (see Appendix \ref{a:implementation-performance}), an inherent property of any system that provides a full transparency log. 
This does mean that as new voters are registered, the size of the log will grow nontrivially. While outside the scope of this paper, future work may consider deleting or archiving records, in accordance with applicable record preservation laws.
\section{Conclusion} 

We have presented \sysname, the first system to bring cryptographic transparency to voter registration systems, and \sysnamepriv, an extension enabling privacy-enhancing execution of cross-database protocols. Our designs are provably secure and verifiable, and our prototypes are efficient and scalable.

Cryptographic measures are no silver bullet for the complex trust issues that are plaguing many electoral systems worldwide~\cite{crisis-of-trust}.  Undoubtedly, increased security and transparency and cryptographically verifiable evidence will be unconvincing to some. 
The current crisis of trust in elections is at least as much as a political problem as a systems one, and as such, must be addressed by both technical and political measures. In this paper, we focus on technical aspects, which we believe are critical to strengthen, even if that will not suffice alone.

We aim to describe
a practical approach that could be fielded in election systems in the near term. 
Based on our informal consultations with current and former election officials and other stakeholders, we designed \sysname modularly to minimize the burden of integration with existing systems. As noted, \sysname does not replace existing voter registration systems. Instead, reminiscent of \emph{end-to-end verifiability} proposals for the casting-and-tallying parts of election systems~\cite{electionguard}, \sysname can be \emph{layered on} to a jurisdiction's existing system to provide extra transparency and verifiability in registration.
Using \sysname or \sysnamepriv should only increase the integrity and confidentiality of a voter registration system, as all data is protected according to jurisdictions' existing access policies. 

We plan to explore piloted adoption in future work, and hope our open-source implementation will promote trust, adoption, and future improvements.


\bibliographystyle{plain}
\bibliography{main}

\appendix
\section{Overview of Consultations}\label{a:interviews-table}
In Table~\ref{tbl:consultations} we present details on the profiles of the (anonymized) stakeholders that were part of our informal consultations described in Section~\ref{sec:voter-reg-background}.

\begin{table*}[!htb]
\centering
\renewcommand{\arraystretch}{1.5}

\begin{tabularx}{\linewidth}{lX}

\hline
\textbf{Primary type of relevant experience} & 
\textbf{Years of election-related experience} \\
\hline
Election official (state) and government & $\geq$ 15 years \\
Industry (election technology) & $\geq$ 20 years \\
Election official (local) and industry (election technology) & $\geq$ 10 years \\
Academic & $\geq$ 20 years \\
Election official (state) and civil society organization & $\geq$ 10 years \\
Industry (election technology) & $\geq$ 5 years \\
Election official (state) & $\geq$ 25 years \\
Election official (state) & $\geq$ 5 years \\
Election official (state) & $\geq$ 5 years \\
Election official (state) and government & $\geq$ 5 years\\
Civil society organization & $\geq$ 15 years \\
Election official (state) & $\geq$ 5 years \\
Election official (local) and government & $\geq$ 20 years \\
Civil society organization & $\geq$ 20 years \\
Lawyer & $\geq$ 20 years \\
Lawyer & $\geq$ 10 years\\
\hline
\end{tabularx}

\caption{\small Brief anonymized profiles of the stakeholder consultations we performed.}

\label{tbl:consultations}
\end{table*}
\section{Additional Security Analysis}
\label{a:more-thms}

\begin{pfsketch}
    \textbf{(Theorem~\ref{thm:public-secrecy})}
    The \emph{privacy} properties of $\reg$ guarantee that the public commitments and update proofs in $\bulletin$ reveal no information about the state of $\reg$. The encryption of the non-public fields guarantees indistinguishability of the published encryptions. Lastly, the fact that each update re-encrypts all fields hides update patterns.
\end{pfsketch}

\begin{pfsketch}
    \textbf{(Theorem~\ref{thm:voter-transparency})}
    Follows from the \emph{soundness} and \emph{completeness} properties of $\reg$, which guarantee that, for any subset of snapshots of the log, a history proof will verify correctly if (correctness) and only if (soundness) it corresponds to the correct data stored in the log.
\end{pfsketch}

\paragraph{Informal transparency theorems (2) and (3).} Informal transparency theorem (2)
follows from the soundness property of $\reg$: update proofs for two consecutive commitments posted in $\bulletin$ only verify correctly if $\reg$ is append only.

Informal transparency theorem (3) is very similar to Theorem~\ref{thm:voter-transparency}, but replacing \Query with \Maintenance: for any third-party and any field they are approved to access, election officials cannot reveal the field in a way that is not consistent with the data audited by voters. As before, the correctness and soundness of $\reg$ guarantees that a third party $\thirdparty$ sees the same content in $\reg$ as the voters. However, this only ensures that $\thirdparty$ has the correct \emph{ciphertexts}, but need not imply correctness of plaintext data: a ciphertext can potentially be decrypted to different payloads under different keys. As such, $\thirdparty$ and a voter can be served different keys, resulting in conflicting views of the data. This is the reason why our construction relies on a key-committing (i.e., binding) encryption scheme, which guarantees that a ciphertext can only decrypt successfully with the correct key, and thus matching ciphertexts imply matching plaintexts. 

\section{Detailed \sysnamepriv Modules}
\label{a:sysnamepriv-modules}

Here, we define the detailed \sysnamepriv modules described in Section \ref{sec:privacy-ehancing-extensions}.

\renewcommand{\algorithmicrequire}{\textbf{Input:}}
\begin{figure}[!htb]
\begin{algorithm}[H]
\caption{\sysnamepriv.\Query}
\begin{algorithmic}
\small
\Require $\textcolor{red}{P}, \id, \{k_{j,i}\}_{j \in [n], i \in [e]}, \big[(\reg_{\id_i}, t_i)\big]_{i \in [e]}, \Pi^{hist}, [\regcom_i]_{i \in [e]}$
\If{$\regverhistory(\id, \big[(\reg_{\id_i}, t_i)\big]_{i \in [e]}, \Pi^{hist}, [\regcom_i]_{i \in [e]}) = 0$}
\State \textbf{abort}
\EndIf
\For{$i \gets 1$ to $e$}
\For{$j \gets 1$ to $n$}
\If{$\public(\voter, \vrdbcol) = 1$} $f_j = \reg_{\id_i}[j]$
\Else
\State \textcolor{red}{$c', c'' = \reg_{\id_i}[j]$}
\State $f_{j} = \Dec_{k_{j,i}}(c')$ \Comment{Abort if $f_j$ is incorrect.}
\State \textcolor{red}{$\hat{c} = \encode({f_j})$} \Comment{Abort if $\hat{c} \neq c''$.}
\EndIf
\EndFor
\EndFor
\State \Return
\end{algorithmic}
\end{algorithm}
\caption{\Query module of \sysnamepriv. The additions to \sysname are marked in \textcolor{red}{red}. Note that checking that the stored encoding matches the encoded plaintext data may require additional information from election officials (e.g., a secret key), which is left implicit in this algorithm.}\label{fig:priv-query-algo}
\vspace{-0.4cm}
\end{figure}


\newcommand{\encodings}{\mathsf{encodings}}
\renewcommand{\algorithmicrequire}{\textbf{Input:}}
\begin{figure}[!htb]
\begin{algorithm}[H]
\caption{\sysnamepriv.\Maintenance}
\begin{algorithmic}
\small
\color{red}
    \Require $P \coloneqq (\encode, \match), \reg^B, \vrdbcols$
    \State $\encodings \gets ()$
    \For{$\id \in \reg^B$}
    \For{$j \gets 1$ to $n$}
    \State $c', c'' = \reg^B_{\id}[j]$
    \State $\encodings \gets c''$
    \EndFor
    \EndFor
    \State $\match(\encodings)$
    \State \Comment{Manually inspect matches, and apply updates as needed.}
    \State \Return
\end{algorithmic}
\end{algorithm}
\caption{The main steps involved in jurisdictions $A$ and $B$ detecting duplicates using \sysnamepriv. This protocol is run by both jurisdictions (shown here for $A$). This module replaces \sysname.\Maintenance.}\label{fig:priv-maintenance-algo}
\vspace{-0.4cm}
\end{figure}

\renewcommand{\algorithmicrequire}{\textbf{Input:}}
\begin{figure}[!htb]
\begin{algorithm}[H]
\caption{\sysnamepriv.\Reg}
\begin{algorithmic}
\small
\Require $\vrdb_{\voter} \coloneqq (f_{1}, ..., f_{n}), \textcolor{red}{P \coloneqq (\encode, \match)}, M, e, $
\State run $\basesys.\Reg(\vrdb_{\voter})$ \Comment{Abort if base system aborts.}
\State $\id_\voter \gets F(K_{id}, \vrdb_{\voter})$
\State $r \gets \bot$
\For{$j \gets$ 1 to $n$}
\If{$\public(\voter, \vrdbcol_j) = 1$}
\State $r \gets r \concat f_j$
\Else \State $k_{j} \gets \KDF(K_{kdf}, \id_\voter \concat \vrdbcol_j \concat e)$ 
\State $r \gets r \concat \Enc_{k_{j}}(f_{j}) \textcolor{red}{\concat \encode(f_{j})}$
\EndIf
\EndFor
\State $M \gets M \concat \add$
\State $\pool$.add($\id_\voter, r$)
\State \Return $\id_\voter$
\end{algorithmic}
\end{algorithm}
\caption{The main steps involved in registering new voters using \sysnamepriv. The additions to \sysname are marked in \textcolor{red}{red}.}\label{fig:priv-register-algo}
\vspace{-0.4cm}
\end{figure}
\clearpage
\section{Implementation Performance}
\label{a:implementation-performance}

The following graph depicts the storage used by our implementation based on the number of records in the log.

\label{storage}
\begin{figure}[!htb]
    \centering
    \includegraphics[width=8cm]{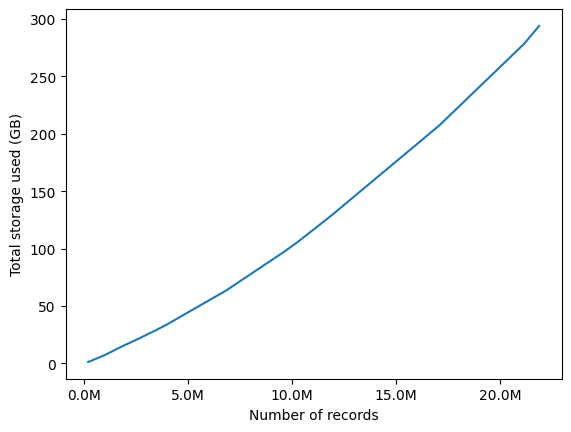}
    \caption{Storage used given number of 1KB records in log.}
    \label{fig:storage}
\end{figure}

\end{document}